\documentclass[conference]{IEEEtran}
\usepackage{amsfonts}
\usepackage{amsmath}

\ifCLASSINFOpdf
  \usepackage[pdftex]{graphicx}
  \graphicspath{{../pdf/}{../jpeg/}}
  \DeclareGraphicsExtensions{.pdf,.jpeg,.png}
\else
\fi
\usepackage{tikz}
\usepackage{textcomp}
\usepackage{hyperref}
\usepackage{lipsum}

\newcommand\copyrighttext{%
  \footnotesize \textcopyright 2017 IEEE. Personal use of this material is permitted.
  Permission from IEEE must be obtained for all other uses, in any current or future
  media, including reprinting/republishing this material for advertising or promotional
  purposes, creating new collective works, for resale or redistribution to servers or
  lists, or reuse of any copyrighted component of this work in other works.
  DOI: \href{https://doi.org/10.1109/ICDMW.2017.8}{10.1109/ICDMW.2017.8}}
\newcommand\copyrightnotice{%
\begin{tikzpicture}[remember picture,overlay]
\node[anchor=south,yshift=10pt] at (current page.south) {\fbox{\parbox{\dimexpr\textwidth-\fboxsep-\fboxrule\relax}{\copyrighttext}}};
\end{tikzpicture}%
}

\begin{document}
\title{Intent-Aware Contextual Recommendation System}

%
\author{\IEEEauthorblockN{Biswarup Bhattacharya\IEEEauthorrefmark{1},
Iftikhar Burhanuddin\IEEEauthorrefmark{2},
Abhilasha Sancheti\IEEEauthorrefmark{3} and
Kushal Satya\IEEEauthorrefmark{4}}
\IEEEauthorblockA{\IEEEauthorrefmark{1}Department of Computer Science, USC Viterbi School of Engineering\\
University of Southern California,
Los Angeles, CA 90089. USA.\\ Email: bbhattac@usc.edu}
\IEEEauthorblockA{\IEEEauthorrefmark{2} \IEEEauthorrefmark{3}Adobe Research\\
\IEEEauthorrefmark{2}Email: burhanud@adobe.com\\
\IEEEauthorrefmark{3}Email: sancheti@adobe.com}
\IEEEauthorblockA{\IEEEauthorrefmark{4}Adobe\\
Email: satya@adobe.com}}

\maketitle
\copyrightnotice

\begin{abstract}
Recommender systems take inputs from user history, use an internal ranking algorithm to generate results and possibly optimize this ranking based on feedback. However, often the recommender system is unaware of the actual intent of the user and simply provides recommendations dynamically without properly understanding the thought process of the user. An intelligent recommender system is not only useful for the user but also for businesses which want to learn the tendencies of their users. Finding out tendencies or intents of a user is a difficult problem to solve.

Keeping this in mind, we sought out to create an intelligent system which will keep track of the user's activity on a web-application as well as determine the intent of the user in each session. We devised a way to encode the user's activity through the sessions. Then, we have represented the information seen by the user in a high dimensional format which is reduced to lower dimensions using tensor factorization techniques. The aspect of \textit{intent} awareness or \textit{intent} scoring is dealt with at this stage. Finally, combining the user activity data with the contextual information gives the recommendation score. The final recommendations are then ranked using some filtering and collaborative recommendation techniques to show the top-k recommendations to the user. A provision for feedback is also envisioned in the current system which informs the model to update the various weights involved in the recommender system architecture. Our overall model aims to combine both frequency-based and context-based recommendation systems and quantify the \textit{intent} of a user to provide better recommendations.

We ran experiments on real-world timestamped user activity data, in the setting of recommending reports to the users of a business analytics tool/application and the results are better than the baselines. We also tuned certain aspects of our model to arrive at optimized results.
\end{abstract}
\IEEEpeerreviewmaketitle

\section{Introduction}
In the business world, showing the advertisement of the ``appropriate" product to the ``appropriate" person at the correct time is the most definite method of attracting customers. This is the general underlying thought behind digital marketing solutions which exist in the current ecosystem. Occasionally, it may be easy to figure out what the user wants but generally it isn't that trivial. Therefore, there is a need for an effective recommendation system which can identify what the user \textit{wants to do} and \textit{needs to do}. 

When we talk about what a user \textit{wants to do}, there are three questions involved. The first question is: Does the user like what he is doing? This is probably a harder question to answer but after analyzing in a web-based situation this reduces to observing things like the frequency of visits, the duration spent by the user on a page, the depth of each visit and so on. Determining the answer to this question at least for a customized scenario is an essential component in improving the recommendations for a user so that the user ultimately starts liking the material shown/suggested to him and he starts spending more time on the webpage. The sweet spot in any recommender system algorithm is to get positive reinforcement which can lead to more revenue for the businesses involved and greater satisfaction of the user. Thus, it is a win-win situation for everyone involved. The second question is: Does the user want to do what he did the last time? For example, an electronic music listener would like electronic music recommendations with high probability. In such situations, we can extrapolate that we can get better results if we suggest them electronic music soundtracks rather than rock music. This leads us on to the third question: Does the user want to do something new that he never did before? This is a very difficult question to answer and much work has been done on providing diversification in recommendation systems to handle problems exactly like this \cite{diversity}. If we can identify what is going on in user's mind then we can design a ``smart" system.

Coming on to the second part of what the user \textit{needs to do}, there are certain aspects involved. A user in a business setting (for eg. an employee using a company's product) generally has an objective provided to him by the administration that he needs to fulfill. Thus, understanding the function of the user and his requirements is an essential component to designing a suitable recommendation system. Speaking from a virtual assistant perspective, if a user is in the sales department of a company then he will want suggestions related to methods of improving sales and may not be bothered about getting computer programming suggestions. However, the programming suggestions may be useful to the technical team who primarily do not have to bother about the sales issues. Thus, even though as a company, both of these two suggestions may be of high relevance to the overall group of ``employees of the company", we see that our recommender system must understand what each user in that company \textit{needs to do} to provide effective recommendations.

A simple frequency-based recommender system essentially remembers the pages visited in the previous sessions by the user and suggests those pages itself. This is a simple algorithm but has obvious issues like not taking into account the content of the page.

An alternate system of recommendations which exists is a content-based recommendation system. Lots of variations exist in literature about this \cite{content1,content2,content3,content4,content5} however, the basic idea remains the same i.e. the system essentially tries to observe the usefulness of a piece of content (or webpage) to a user and thus recommend it to a user. Usually, there is some sort of relevance ``score" involved which can range from simple methods like cosine similarity to more complex algorithms.

The novelty of our work lies in the focus on combining these two aspects of recommender systems in the hope of providing better recommendations. We propose an innovative method of encapsulating the user intent so as to increase the relevance of the recommendations shown. In complex systems, there are a vast array of parameters which can be changed, thus leading to thousands of possible recommendations which can be suggested to the user. Overall, in this paper, we suggest a methodology to rank the content and combine this ranking with historical browsing knowledge to arrive at a new recommendation score. We conduct experiments on real data for recommending reports to users to demonstrate the effectiveness of our approach.

\section{Related Work}
Firstly, we have a content-based recommendation system which recommends particular kinds of items to the users. This consists of the recommendation system used by Netflix \cite{netflix}, traditional music recommendation systems \cite{music} and video recommendation by YouTube \cite{youtube}. These kinds of recommendations are driven by the content of the items to be recommended (i.e. description of the music, movie title, genre etc.) and a profile of the user's preference based on the historical information. A way of combining the global popularity and the user preference of videos to recommend the next N videos recommendation on YouTube is described in \cite{youtube}. Various methodologies like use of latent factors to model the relationship between items and users and to determine the rating of videos by the neighboring videos seen by the user are discussed in \cite{netflix}.

Other kinds of systems deal with a Markov-model approach. This consists of models which study item similarity to understand sequential patterns and recommend next set of items based on the previous set \cite{basket}. User's preference for different items can be modeled using a Markov chain. Therefore a transition matrix is estimated that gives the probability of buying an item based on the previous purchases of the user. Tensor decomposition methods can be used to promote collaborative filtering by combining transition matrices of similar users. However these models lack the knowledge of contextual signals from the user. Then, we have  context-based recommendation systems. These systems represent context(such as time, place) either explicitly through various external signals \cite{cb,mc} or implicitly using the technique of exploration and exploitation \cite{cr}. Tensor decomposition methods can be extended to combine content, context and collaborative filtering as in \cite{cb}. Bandit's approach can be used to generate context aware online recommendations. Instead of modeling context explicitly through external signals, \cite{cr} uses user activity or user feedback to model the context and hence gives the most suitable recommendation in that context. These systems lack sequential evolution of the state and they do not consider the final intent of the user. 

Finally, some literature exists on tracking user intent by analyzing context. Contextual Markov chains \cite{mm} and nowcasting models \cite{context} have been studied to track user intent through the context. However context modeling in these cases are limited to consumer applications. Context and intent in enterprise applications have to be dealt with differently and modeled in a way relevant for the enterprise users.

The motivation behind our work comes from the intent tracking with sequential evolution as described in \cite{cb}. We have redesigned the algorithm combined with other techniques to suit our needs.

\section{Problem Definition}
In this section, we formally define the problem we target to solve in this paper.

There are certain special business intelligence tasks involved which require the user to go through the content of multiple reports to arrive at conclusions. Often, specialized business intelligence tools are used for this purpose. Thus, the user has an end goal in mind which is of prime importance to the recommendation system. We refer to this goal as the intent of the user which in our scenario is a report. This is a very difficult task to solve given the current literature. Our system aims to provide recommendations of this manner where the system tries to understand the end goal of the user thereby assisting the user in coming to conclusions faster. This also helps in reducing human error as the system essentially tries to understand the quality of the webpage content (which in the normal case the user tries to utilize manually) and suggest webpages (reports in our scenerio) accordingly. 

Also, in our case, a new user of a tool may be overwhelmed by the number of features provided by it. There is not much history involved for a new user so recommendations are not ``correct" or appropriate most of the times. This user's needs have to be understood over a certain period of time. Thus, in this case, a collaborative approach makes sense. There has been some work done in case of expert recommendations in literature \cite{expert}. Our system utilizes a collaborative/group-based model to suggest recommendations in order to address this issue.

In this paper, we aim to solve the following questions: \begin{itemize}
\item Predict user \textit{intent} which is the end goal or report in our scenario, from context and frequency
\item Determine the relationships between various features in the content of a webpage
\item Determine the right content, data and representation based on the type and expertise of the user
\item Find the most suitable recommendation scoring system
\end{itemize}

We will propose a model that is able to address the above questions. We will refer recommendation items as reports and users as analysts in the paper, but the approach works with any type of items with parse-able information (e.g. webpages, products, movies, music etc.).

\section{System Design}
In this section, we describe our approach in detail. We first describe the graphical and contextual models that predict the relevance of items to users (Sections IV-A,B,C,D), then we describe how we can incorporate these two models together to generate the recommendations (Section IV-E,F). Here, items refer to the reports seen by the users. Each report is a dimension which gives the value of a metric for each of the dimension elements. Metric is a continuous variable, dimension is a categorical variable and dimension element is a value that a dimension takes. For example, for a report ``Page Views" which consists of the page views of individual pages in a website, the dimension elements can be \{``homepage'', ``sales page'', ``contacts page''\} and so on and the metric for these elements is ``number of views''. Time-series reports capture the variation of a metric over time for a particular dimension element while histogram reports consist of histograms of metrics over different dimension elements. Finally, we briefly discuss about feedback in this scenario (Section IV-G).

We propose a solution to predicting user intent based upon current context and combining it with a frequency based graphical model to provide relevant recommendations to the user. We have defined the current context as the current report being looked at by the analyst. Our solution approach has 3 main steps:
\begin{enumerate}
\item Modeling browsing patterns through a user navigation graph (Markov Model) where each node is uniquely represented.
\item Modeling context of the user as a matrix of context features versus time and predicting user intent from latent factors which are obtained from current context.
\item Combining the Markov Model and the intent scores to give relevant recommendations to the user.
\end{enumerate}

\subsection{User Navigation Graph}
\begin{figure}[h]
\caption{Frequency Model using hit data (Section V-A) \label{fig:frequency}}
\centering
\includegraphics[height=5in,width=3in, keepaspectratio]{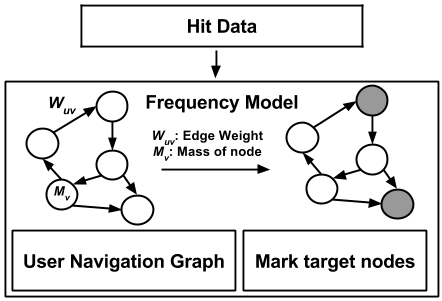} 
\end{figure}

The first step of the algorithm is to generate a navigation graph for each user. We need to learn the reports (webpages) that each of the users visit so that we can build our set of reports from which to recommend. We do this by encoding additional information while building this \textit{graph} of reports for every user. This forms the \textit{frequency-based} recommendation aspect of our algorithm. The frequency model is illustrated in Figure \ref{fig:frequency} and explained below.

A user navigation graph is a weighted directed graph with each node signifying a unique report seen by the user and each directed edge between node $u$ and $v$ signifying that the user transitioned between node $u$ and $v$ with probability greater than $0$. This user graph can be thought of as a Markov model since the outgoing transition probabilities for each node sums up to $1$.

\subsection{Determining the Target nodes}
The second step of the algorithm is to determine the target/intent nodes. The motivation behind this is that every user has some intent in his mind as was discussed before. The target nodes are denoted by setting the \textit{target} attribute of the node to $1$. Target nodes refer to the possible candidates for intent which is the end of analysis or a milestone. The target nodes were arrived by analyzing the sequences of nodes visited by the user in a session and using the following heuristics. If the node in question has in-degree greater than or equal the mean of the in-degrees of all nodes, then the \textit{target} attribute is set to $1$. This algorithm has yielded at least $1$ target node even in the smallest of graphs particularly due to the inherent property of the statistical mean which ensures that at least $1$ node will be set as a \textit{target}.

\subsection{Context Model}

The next step of the algorithm is to model the user context. This forms the basis of the \textit{context-based} recommendations as was discussed before. This combined with the \textit{frequency} information encoded in the user navigation graph will give the final recommendation score. This combination is the novel step which is behind the improvement in performance. Section V-D gives the evidence of this improvement.
\begin{figure}[h]
\caption{Building the context tensor}
\centering
\includegraphics[height=5in,width=3in, keepaspectratio]{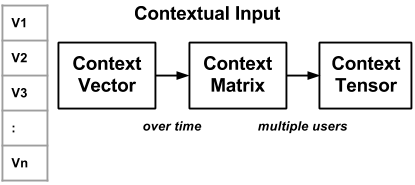} 
\end{figure}

There are mostly two kinds of reports seen by the users: one is the report in the form of histograms and the another is in the form of time series. To model context from the current report (that the user is seeing), we extract some relevant features as described below. This of course varies from domain to domain based on implementation.

For each time-series report we extract the following $6$ features \cite{timeseries}: aggregate value, maximum and minimum value of the time series, location of maximum and minimum observation, longest positive and negative runs, length of time series \& average absolute change in consecutive observation. These features are calculated for each metric and dimension element pair. Alternatively, for each histogram report only the aggregate value as a feature is populated and rest of the $5$ values in the vector are set to $0$. From all the reports seen by a given user we form descriptions for building the context vector in which we specify a place for six different values for all the reports seen. So at each time instant, the corresponding vector is populated according to the type of the report. By combining context vectors at different time instants, the context matrix is formed for all the users.

The users of the tool are analyzed and clustered into 4 categories depending on their exposure and competence level of using the tool using k-means clustering algorithm. This is done for the purpose of group-based recommendations which is discussed in Section IV-F. The users are categorized depending upon their usage and amount of activity on the tool. Activity here implies duration of browsing, number of redirects, reports seen.  

After clustering, the user matrix is combined for users within a cluster to form the context tensor. For a given user we have a set of different dimension elements over which metric is calculated. The cardinality of this set for each metric $m$ is denoted as $d_m$. We define $D_{u} = \sum_{m=1}^{M_{u}} d_{m}$ where $M_{u}$ is the cardinality of the set of all metrics seen by the user. For each report, we have $6$ observable features. Thus the dimension of the context vector is $6 \times D_{u}$, which we denote as $N^u$. We populate all the $6$ values for a time-series report and only the aggregate value is populated for a histogram report. All the remaining positions are given a value  of zero indicating they are not present in the current report. The aggregation of many such context vectors over the number of reports seen form the context matrix. The dimension of the context matrix is hence $N^{u} \times T$, where $N^{u}$ signifies the context variables which varies from user to user and $T$ signifies the number of reports seen. We form the context tensor consisting of context matrices of $U$ users from the same cluster. The method to generate the context tensor is illustrated in Figure 2.

For each user we repeat the reports seen, such that the total number of reports seen is equal to the maximum number of reports seen by any user. Since the tensor is highly sparse and the number of context features varies from user to user, PARAFAC2 tensor decomposition is used to obtain latent factors for each report seen by the user \cite{context,parafac2}.

\subsubsection{PARAFAC2 Decomposition}
For a three-way tensor i.e. $N^{u}\times T \times U$, PARAFAC2 decomposition requires only two out of the three modes ($N^{u}$, $T$ and $U$) to have uniform sizes, which in our scenario are the time ($T$) and user modes ($U$), while the third mode $N^{u}$ can be of different sizes. 
The decomposition is equivalent to solving the following optimization problem: \begin{equation} {(\mathbf{\tilde{F}} ,\mathbf{\tilde{\Lambda}} ^{u})}_{u = 1,2,...|U|} = \min_{\mathbf{F},\mathbf{\Lambda}^{u}} \sum_{u=1}^{|U|}{||\mathbf{X}^{u}- \mathbf{\Lambda}^{u}\mathbf{F}||^{2}}_{F} \end{equation}
where \textit{F} is the Frobenius norm.
The latent factor space dimensionality is $R\times T$. The value of $T$ is discussed above. The value of $R$ is a constant fixed by us for design purposes. It is typically a value much smaller than the number of users ($|U|$), which leads to the reduction of the dimensionality of the original context tensor ($X$). After decomposition, the panel for the $u^{th}$ user is approximated by 
\begin{equation} \mathbf{X}^{u} \approx \mathbf{G}^{u}\mathbf{HS}^{u}\mathbf{V}' \end{equation}
where $\mathbf{G}^{u} \in \mathbb{R}^{N^{u} \times R}$ is an orthonormal matrix, $\mathbf{H} \in \mathbb{R}^{R \times R}$ is a matrix invariant to $u$, $\mathbf{S}^{u} \in \mathbb{R}^{R \times R}$ is a diagonal matrix and $\mathbf{V} \in \mathbb{R}^{T \times R}$ is the matrix containing the collaborative latent factors at $T$ time instances.
The latent factors are a lower dimensional representation of reports. For the $u^{th}$ user, the initially estimated latent factors are \begin{equation} \mathbf{\tilde{F}}^{u} = \mathbf{\tilde{F}} = \mathbf{V}' \end{equation}
and the factor loading matrix is estimated by \begin{equation} \mathbf{\hat{\Lambda}}^{u} =  \mathbf{G}^{u}\mathbf{HS}^{u} \end{equation}

\subsubsection{Kalman Filter} 
Kalman Filter is used to enforce the dynamics and sequential correlations in the latent factors. Let the  a priori and a posteriori latent factors and error covariance matrices at time step t be $\mathbf{\tilde{f}}_t, \mathbf{\hat{f}}_t,  \mathbf{\tilde{P}}_t^u$ and $\mathbf{\hat{P}}_t^u,$ respectively. $\mathbf{A}^u$ represents the transition matrix. In the time update (prediction) step, the a priori factors for the next time step are computed by 
\begin{equation} \mathbf{\tilde{f}}_t= \mathbf{A}^u\mathbf{\hat{f}}_{t-1} + \boldsymbol{\omega}_t^u \end{equation} 
and the priori error covariance by \begin{equation} \mathbf{\tilde{P}}_t^u = \mathbf{A}^u \mathbf{\tilde{P}}_{t-1}^u \mathbf{A}^{u'} + \mathbf{Q}^u \end{equation}
In the measurement update (correction) step, the Kalman gain $\mathbf{K}^u_t $ equals
\begin{equation} \mathbf{K}_t^u = \mathbf{\tilde{P}}_t^u \mathbf{\Lambda}^{u′} (\mathbf{\Lambda}^u\mathbf{ \tilde{P}}_t^u \mathbf{\Lambda}^{u'} + \mathbf{\Psi}^u)^{-1} \end{equation}
where $\mathbf{\Psi}$ is the covariance matrix \cite{context}. With the Kalman gain, the a priori latent factors are corrected by available contextual signals (missing signals are assigned a very large variance), and the a posteriori factors equals
\begin{equation} \mathbf{\hat{f}}_t = \mathbf{\tilde{f}}_t+ \mathbf{K}_t^u\mathbf{(x}_t^u - \mathbf{\hat{A}}\mathbf{\tilde{f}}_{t}) \end{equation}
The a posteriori error covariance for next time step equals
\begin{equation} \mathbf{\hat{P}}_t^u = (\mathbf{I} - \mathbf{K}_t^u\mathbf{\Lambda^u})\mathbf{\tilde{P}}_t^u \end{equation}
$\mathbf{I}$ is the identity matrix
The value of R was iterated over multiple values to arrive at optimal values so that latent factors can be evolved faster \cite{context}.

\subsection{Ranking of Latent Factors}

To distinguish between the finally evolved latent factors (using Kalman filtering) with and without intent $\textit{I}$ for each user, the pair-wise learning-to-rank method of RankSVM \cite{rank} was used.
\begin{figure}[h]
\caption{Context Model to get relevance scores \label{fig:context}}
\centering
\includegraphics[height=3in,width=3in, keepaspectratio]{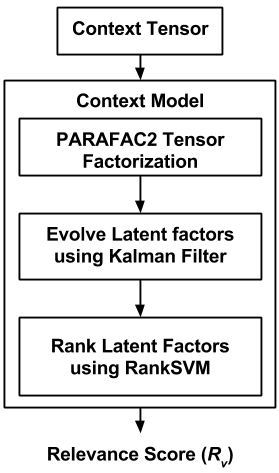} 
\end{figure}

Based on the training data we have the set $\mathbf{R}_1$ which consists of all the latent factors $\mathbf{f}_i $ such that the user eventually ends up at target node $\textit{I}$ in the session. Similarly we have the set  $\mathbf{R}_2$ which consists of all the latent factors $\mathbf{f}_j$ such that the user does not eventually end up at target node $\textit{I}$ in the session i.e. the user reaches a target node other than the intent $\textit{I}$.

We define $g(\textit{I},\mathbf{f}_i) = 1$ if $\mathbf{f}_i$ belongs to the set $\mathbf{R_1}$ otherwise $g(\textit{I},\mathbf{f}_i) = 0$ if $\mathbf{f}_i$ belongs to the set $\mathbf{R_2}$. So, this notation induces a partial ranking of latent factors for intent $\textit{I}$ such that $\mathbf{f}_i$ is preferred over $\mathbf{f}_j$ for intent $\textit{I}$. P is a set as defined below. 
$P = \{ (i,j) : \mathbf{f}_i \in \mathbf{R}_1 \hspace{0.5em} and \hspace{0.5em} \mathbf{f}_j \in \mathbf{R}_2 \}
$

The objective of the RankSVM is to learn the ranking function $g(\textit{I},\mathbf{f}_i)$ such that $g(\textit{I},\mathbf{f}_i) > g(\textit{I},\mathbf{f}_j) $ for all $\mathbf{f}_i \in \mathbf{R_1}$ and $\mathbf{f}_j \in \mathbf{R_2}$. Thus, $g(\textit{I},\mathbf{f}_i)$ can be defined as 
\begin{equation} g(\textit{I},\mathbf{f}_i) = < \vec{w},\mathbf{f}_i > 
\end{equation}

Following a large margin approach leads us to the optimization problem:
\begin{equation} \min_{ \vec{w},\epsilon_{ij} \geq 0 } <\vec{w},\vec{w}> + \lambda \sum_{ij} \epsilon_{ij} \end{equation}
$s.t. \hspace{0.3em}
\forall (i,j) \in P, <\vec{w},\mathbf{f}_i>$  $\geq$  $<\vec{w},\mathbf{f}_j> + \hspace{0.3em} 1 - \epsilon_{ij} 
$, where $\lambda>0$ determines the trade-off between margin maximization and error minimization. The latter is the sum of individual losses $\epsilon_{ij}$ and constitutes an upper bound on the 0/1-loss of mistaken preference relations. The constraints
enforce $< \vec{w}$, $\mathbf{f}_i >$ is greater than $< \vec{w}$, $\mathbf{f}_j$ $>$  whenever possible and penalize violations
thereof. Once optimal $\vec{w}$ has been learned they can be used to induce ranking of new latent factors for each intent.

From the ranking function we wish to obtain the intent score ($S$) such that scores are normalized to be between $0$ and $1$. Here, $\vec{w} \cdot \vec{\mathbf{f}}$ represents the distance between the training point and plane separating positive training and negative training points. The latent factors are normalized to have norm $1$. Now since the RankSVM measures the distance between the hyperplane and the set of points, the distance value lies between $-4$ and $4$ (since the difference of latent factors have norm less than $2$ after normalization).

For every test data, we calculate the latent factor ($\vec{\mathbf{f}}$) for the report seen and take its dot product with the weights for each intent which finally gives us the intent score ($S$) , normalized to be between $0$ and $1$:
\begin{equation}S_{\textit{I}}(\vec{\mathbf{f}}) = \frac{(4+\vec{w}_{\textit{I}}\cdot\vec{\mathbf{f}})}{8}
\end{equation}

The method to get the relevance score through the context model is illustrated in Figure \ref{fig:context}.

\subsection{Recommendation Scoring}
\begin{figure}[h]
\caption{The recommendation system with feedback \label{fig:recscore}}
\centering
\includegraphics[height=4in,width=3in, keepaspectratio]{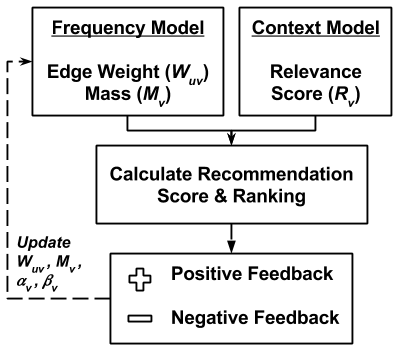} 
\end{figure}

The next step of the algorithm is to determine the recommendation scores for the candidate recommendable nodes. This is the step which puts together \textit{frequency-based} and \textit{context-based} recommendations. According to our hypothesis, our aim was to prove that combining both of these can actually improve recommendations.

This step is responsible for finally combining the intent scores, edge weights ($W_{uv}$) and the masses ($M_{v}$). The formula that we are using for scoring recommendation ($K_{uv}$) is as follows:
\begin{equation}
K_{uv}  = (\alpha_{v} \times W_{uv} \times R_{v}) + (\beta_{v} \times M_{v})
\end{equation}

The components of the recommendation score are as follows:
\begin{itemize}
\item $W_{uv}$: This is the historical probability that the user goes from current node, $u$ to a recommendable node, $v$. This probability information is stored in the graph itself as the edge weight.
\item $R_{v}$: This is the relevance score of the recommendable node, $v$. This score is essentially a function of the intent scores and the probabilistic distance from the node $v$ to the set of target (intent) nodes (the function definitions Max-IxD and Dot-IxD are discussed in Section V-D). Using single source Dijkstra shortest path algorithm, we calculate the distance between the node $v$ and the target nodes in the graph. In our case as the edge weights of the graph are not the distance between two nodes but the probabilities of transition, probabilities are to be multiplied when traversing through the graph. Keeping this in mind, we converted the edge weights to their negative natural logarithms (as natural logarithm of probabilities which are values less than $1$ give negative values, a negative sign is added to make the weights positive). Thus, adding logarithms corresponds to multiplication of the probabilities. Hence, with this modification, single source Dijkstra shortest path algorithm was applied to find these shortest paths. Finally, for the node $v$, a list of targets and the probability of reaching those targets from the node $v$ based on the user's historical user navigation graph was calculated.
\item $M_{v}$: It is the fraction of the total browsing time of the user that he spent on the node $v$. The word ``mass" is being used to denote the notion of weighing down i.e. if the mass of a node is relatively high, the user is weighed down in that node as he spends more time there compared to the other nodes.
\item $\alpha_{v}$, $\beta_{v}$: These are the feedback factors of node \textit{v}. These factors are initialized with $1.0$. However, they change on receiving feedback.
\end{itemize}

The recommendation scoring technique is illustrated in Figure \ref{fig:recscore}. The recommendation score is generated for all nodes \textit{v} connected to the current node \textit{u}, as well as for all nodes \textit{w} connected to each of the nodes \textit{v}. Hence, essentially, 1-step and 2-step distanced nodes from $u$ get a score. This calculation is done for all the user graphs available irrespective of the current user.

The reasoning behind the formula is henceforth explained. The $R_{v}$ gives the true score of the relation between the current user context and intent. This score incorporates the report data and other stimuli.

The $R_{v}$ calculation results in the scoring of the intents but it does not really deal completely with the actual history involving the nodes to be recommended. It is more of a score which assures that from node $v$ a certain set of intents can be reached or, should be reached (goals or reports which are required to complete an analysis). However, the past user behavior might indicate other tendencies of the users. Historical analysis might reveal that the user is always interested in certain kinds of reports which may not always be encoded by the contextual information. For handling such kind of cases, multiplying the edge weight with the relevance score makes logical sense as the edge weight itself can act like a weighing factor to the relevance score. Thus, a node with high $R_{v}$ which a user rarely visits may not be shown as the top recommendation due to the fact that the user is not likely to find it useful as per their browsing history. Thus, in most cases, it is expected that the edge weight as well as relevance should be positively correlated and the multiplication would not affect the ordering. However, for cases where the correlation between the edge weight and the relevance is negative then the multiplication plays a role in trading off each value with each other and arriving at a more realistic prediction. Finally the mass is added as an offset to the recommendation score. It is generally a small factor which signifies the importance of a node. In cases where the $R_{v} \times W_{uv}$ is similar, the mass will determine the ranking as the actual time spent by the user historically must be factored in. In other cases, based on empirical analysis, the mass does not play a deciding role as such. However, the mass attribute can be used finally to break ties if there are any (discussed further in Section V-B).

\subsection{Group-based Recommendations}
Apart from recommending reports to a user from that user's graph we also recommended reports from other user graphs or user groups. This step is especially useful for new or novice users who do not have much personal history with the tool. For such users, sourcing recommendations from an ``experienced" group is a good idea and it has been shown to work.

First, to identify the users, we selected user groups based on clusters. For new or inexperienced users, we gave recommendations from clusters which contained more experienced users. Secondly, after identifying the relevant clusters we identified the set of users who have the same report/node in their graph and at least one of the target node. Finally, for the selected users, we calculated the recommendation score in a similar manner as we calculated above but with some slight variations. Intent scores ($R_{v}$) are the same as the intent scores of the original user (the user to whom the recommendations are shown). Edge weights ($W_{uv}$) and mass ($M_{v}$) are taken from the selected users' (group) graphs. Therefore, the notion of ``distance" as discussed before is also according to those selected users' (group) navigation graph.

We calculated recommendation scores only for those nodes from other users which were not present in the original user graph i.e. nodes novel to the original users. These recommendation scores are combined with the recommendation scores of original user to give the final set of recommendations. In the next section we will see that feedback can be used to inform the model and provide better recommendations.

\subsection{Feedback}
The final step of the algorithm is to gather implicit/explicit feedback and update the system. Feedback is a very important requirement in any ranking system as it informs the algorithm about its own correctness and rectification can be done in the subsequent calculations. We defined feedback in the following situations:
\begin{itemize}
\item Explicit positive feedback: If a user clicks on one of the recommendations shown in the list.
\item Explicit negative feedback: If a user deletes one of the recommendations shown in the list. 
\item Implicit positive feedback: If a user does not click on any of the recommendations shown in the list, however navigates to a page which existed in that list. This is particularly an important piece of information as this addresses many of the problems existing in today's recommendation system. The recommender intelligently identifies the actions of the user and updates accordingly.
\item Implicit negative feedback: If a user does not click on any of the recommendations shown in the list and navigates to a page which did not exist in that list. Thus, even though the user did not delete any recommendation, this can be taken to be a weak signal of negative feedback.
\end{itemize}

This leads to a change in the $W_{uv}$, $M_{v}$, $\alpha_{v}$ and $\beta_{v}$ values which were introduced before. The appropriate values can be updated on the basis of mathematical optimization methods as well as empirical observations. Also, the updating algorithm must be slightly different for giving feedback on 2-step nodes compared to 1-step recommendations. A simulation system is planned to evaluate the feedback mechanism.

\section{Experiments}
In this section we will showcase the performance improvements of our model compared to baseline models. We will first describe the dataset, the algorithm setup with respect to our case and then evaluate our model with other baselines.

We also conduct several case studies to show which
recommendation score is the most appropriate.

\subsection{Dataset}
We used real-world hit data. A hit corresponds to a report access. The hit data essentially consisted of time-stamped user activity tracking. Data was filtered to give relevant information like timestamp, report requested, URL, metrics seen, and so on. The data also has information related to the session which is used to sessionize it for context modeling.

We used 10 days hit-level data. This data is taken across different companies so the number of users varies depending upon the company. The number of reports visited by each user also varies depending upon his tasks. The data contained information about the user as well as the reports visited by each user at each timestamp. The data was finally sessionized to understand and segregate the transitions. We trained our model on 70\% of the data (first 7 days) and tested on the remaining 30\% (last 3 days) data. Further work may be conducted by using alternative training-testing methods like interleaved training, increasing the percentage of the training set, increasing the amount of overall data and so on.

\subsection{Ranking of Recommendations}
The recommendations generated using the methodology described in Section IV are stored along with the information such as the value of recommendation score, $R_{v}$, $\alpha_{v}$, $\beta_{v}$, $M_{v}$, $W_{uv}$ and whether the node is actually the 1-step or 2-step recommendation. Now, these generated recommendations are ranked in the following order of preference:  
\begin{enumerate}
\item $K_{uv}$ (Descending): This is the recommendation score obtained by using Equation (13). It is highly unlikely that two nodes will have the same recommendation score. 
\item Collaborative or not (Preferring the recommendations got from the current user graph over the recommendations got from the other user graphs)
\item $R_{v}$ (Descending): Relevance score (Section IV-E).
\item $W_{uv}$ (Descending): If a node $v$ is not adjacent to $u$ then shortest path weight which is the product of probabilities on the path is taken.
\item $M_{v}$ (Descending): ``Mass'' of the node (Section IV-E).
\end{enumerate}

\subsection{Baselines}
We tested our model against the baselines described below:
\begin{enumerate}
\item \textbf{Frequency}: Recommendation based upon the probabilistic graphical model i.e. based upon edge weights ($W_{uv}$) in the user navigation graph.
\item \textbf{Mass} ($M_{v}$): Recommendation based upon the average time spent on the reports seen.
\item \textbf{Context}: Recommendation based upon intent scores obtained from the current context of the user.
\item \textbf{Tensor Factorization}: Recommendation based upon obtaining latent factors only from PARAFAC2 tensor decomposition (without Kalman Filter regularization).
\end{enumerate}

\subsection{Results}

\begin{table}
\centering
\caption{Comparison of proposed system with baselines \& different $R_{v}$}
\begin{tabular}{|c|c|c|c|c| } \hline
\textbf{Method} & \textbf{NDCG} & \textbf{Precision} & \textbf{Recall} & \textbf{w-AUC} \\ \hline
\textbf{Mass} & 0.4297 & 0.0885 & 0.6181 & 0.5867\\ \hline
\textbf{Frequency} & 0.4744 & 0.0853 & 0.5866 & 0.5866\\
\hline
\textbf{Context Based} & 0.5228 & 0.1003 & 0.6730 & 0.7226 \\
\hline
\textbf{PARAFAC2 Model} & 0.5470 & 0.0987 & 0.6343 & 0.6250\\
\hline
\textbf{Max-IxD} & 0.4908 & 0.1006 & 0.6753 & 0.6858 \\
\hline 
\textbf{Dot-IxD} & 0.5274 & 0.1021 & 0.6911 & 0.7165 \\
\hline
\textbf{Max-I} & 0.4736 & 0.0969 & 0.6390 & 0.6555\\
\hline
\textbf{Sum-I (proposed)} & \textbf{0.5706} & \textbf{0.1006} & \textbf{0.6753} & \textbf{0.7239}\\ \hline\end{tabular}
\end{table}

We tested our model on the $4$ metrics: NDCG (Normalized Discounted Cumulative Gain), Precision, Recall and w-AUC (weighted average Area Under the Curve). The results shown below are for the users seeing considerable number of unique reports. The results for performance of our algorithm against other baselines are summarized in Table $1$.

It is seen that the proposed model i.e. the Markov model with context modeling using PARAFAC2 and Kalman Filtering works better than the other models with respect to the metrics used. The precision is near $0.1$ because we set an upper limit to the number of recommendations to be generated to be $10$ and for every recommended list, there could be only $1$ correct recommendation as testing was done by simply checking the immediate next report access. The number of reports seen/accessible (universe of reports) is a huge number thus driving down overall results in case of precision. However, it is clear from comparison that our algorithm indeed performs better than baselines with respect to each metric.

We have also varied the recommendation scoring formula. In this case, the general formula structure remains the same as Eqn. (13), but the method of calculation of $R_{v}$ changes. The key for the formulae are as follows:
\begin{itemize}
\item \textbf{Max-IxD}: $R_{v}$ = Maximum of (intent scores * distance to target node)
\item \textbf{Dot-IxD}: $R_{v}$ = Dot product between intent score and distances of node to those intents
\item \textbf{Max-I}: $R_{v}$ = Maximum of intent scores
\item \textbf{Sum-I}: $R_{v}$ = Sum of intent scores (proposed)
\end{itemize}

It is observed that Dot-IxD performs better than other methods with Precision and Recall. The intuition behind Max-IxD is to extract the information of the most relevant and visited nodes among the options. Similarly, the intuition behind Dot-IxD is to get a single score which captures not only the maximum but encodes the information of all possible intent scores and the distances. The proposed model in Table $1$ uses the $R_{v}$ as described in the Sum-I method. The intuition behind proposing Sum-I is the fact that it considers the relevance scores fully without any discount and does not lead to any bias. As we are already considering the probabilities of reaching a node through the user navigation graph ($W_{uv}$), it is not necessary to encode the distance information again in $R_{v}$.

\section{Conclusion}
In this paper we have proposed a novel recommendation system combining a probabilistic graphical model with tensor factorization and Kalman filtering. Though some steps like tensor factorization are computationally heavy, they are to be conducted after long periods (once a day). The updating of latent factors through Kalman filter is a fast process thus making the system suitable for usage in real-time. From the results, we are able to justify our hypothesis that combining \textit{frequency} and \textit{context-based} recommendations should lead to better results. Further analysis maybe conducted by tuning the parameters and incorporating the feedback system which should lead to improved results.

\section*{Acknowledgment}
This work was done during an internship at Adobe Research BigData Experience Lab, Bangalore, India in summer 2016. The authors would like to thank Adobe Research for their support.

\bibliographystyle{IEEEtran}
\bibliography{biblio}

\end{document}